\documentclass[
 aps, pra,
 amsmath,amssymb,
 11pt,
 final,
tightenlines,
 twoside,
 twocolumn,
 nofloats,
nofootinbib,
 superscriptaddress,
showkeys,
showkeywords,
onecolumn
 ]
{revtex4-2}

\usepackage[T1]{fontenc}
\usepackage[utf8x]{inputenc}
\usepackage[english]{babel}
\usepackage{graphicx}
\usepackage{dcolumn}
\usepackage{bm}
\usepackage{longtable}

\input{maik.rty}

\newcommand{\kms}{\,km\,s$^{-1}$} 

\setcitestyle{authoryear,round}
\def\saoname{Special Astrophysical Observatory,  Russian Academy of Sciences,
              Nizhnii Arkhyz, 369167 Russia}

\def\squareforqed{\hbox{\rlap{$\sqcap$}$\sqcup$}}

\def\sq{\ifmmode\squareforqed\else{\unskip\nobreak\hfil
\penalty50\hskip1em\null\nobreak\hfil\squareforqed
\parfillskip=0pt\finalhyphendemerits=0\endgraf}\fi}

\def\utw{\smash{\rlap{\lower5pt\hbox{$\sim$}}}}

\def\udtw{\smash{\rlap{\lower6pt\hbox{$\approx$}}}}

\def\diameter{{\ifmmode\mathchoice
{\ooalign{\hfil\hbox{$\displaystyle/$}\hfil\crcr
{\hbox{$\displaystyle\mathchar"20D$}}}}
{\ooalign{\hfil\hbox{$\textstyle/$}\hfil\crcr
{\hbox{$\textstyle\mathchar"20D$}}}}
{\ooalign{\hfil\hbox{$\scriptstyle/$}\hfil\crcr
{\hbox{$\scriptstyle\mathchar"20D$}}}}
{\ooalign{\hfil\hbox{$\scriptscriptstyle/$}\hfil\crcr
{\hbox{$\scriptscriptstyle\mathchar"20D$}}}}
\else{\ooalign{\hfil/\hfil\crcr\mathhexbox20D}}%
\fi}}

\newcommand{\aap}{Astron. and Astrophys. }

\newcommand{\aaps}{Astron. and Astrophys. Suppl. }

\newcommand{\aj}{Astron.~J. }
\renewcommand{\apj}{Astrophys.~J. }

\newcommand{\mnras}{Monthly Notices Royal Astron. Soc. }

\newcommand{\pasp}{Publ. Astron. Soc. Pacific }

\begin{document}

\selectlanguage{english}



\title{On the intrinsic rest wavelengths of diffuse interstellar bands}

\author{\firstname{Gazinur A.}~\surname{Galazutdinov}}
\email{runizag@gmail.com}
\affiliation{Federal State Budget Scientific Institution Crimean Astrophysical Observatory of RAS, Nauchny 298409, Crimea}
\affiliation{\saoname}
\author{\firstname{Elena}~\surname{Babina}}
 \email{helenka_truth@mail.ru}
\affiliation{Federal State Budget Scientific Institution Crimean Astrophysical Observatory of RAS, Nauchny 298409, Crimea}

\begin{abstract}
The rest wavelengths of diffuse interstellar bands (DIBs) are the parameters of fundamental importance owing to the lack of unambiguous identification for these mysterious features.
Usually the wavelengths of DIBs are estimated using known interstellar atomic or molecular lines serving to shift the spectrum to the rest wavelengths velocity scale.
However, the latter may not share in fact the same parts of a cloud which carriers of diffuse bands occupy.
Here we argue that the narrowest known diffuse interstellar band 6196~\AA\ is the best reference feature for building the "interstellar"\, wavelength scale.
Also, we offer the geometrical center of gravity (the effective wavelength) of diffuse bands as a rest wavelength measurer for these generally asymmetric features.
The exception is DIB 6196: its symmetrical (bottom) part must be used as a reference for the interstellar wavelength scale, and its center of gravity
serves to study the variability of the feature visible at the upper part of its profile.
We assessed the magnitude of variation of the center of gravity of diffuse bands at 5780, 5797, 6284 and 7224~\AA\ measured in 41 lines of sight in the broad range of interstellar reddening (E$_{B-V}$ varies in the range 0.13 -- 1.06 stellar magnitudes) with the lack of evident Doppler-split in profiles of interstellar atomic/molecular lines.
We demonstrated that diffuse bands show the gradual broadening of their profile widths, accompanied with red-shift of the center of gravity, i.e. the red wing of the profiles is most variable part of the profiles.
To estimate the width of diffuse bands we offer to apply a parameter "effective width"\, W$_{eff}$, which is the relation of the equivalent width (EW) to the depth of the feature. In contrast to habitual half width at half maximum (FWHM), the parameter originally introduced by us in 2008 (Galazutdinov, LoCurto, Kre{\l}owski) is not sensitive to the profile shape irregularities. W$_{eff}$ provides lower uncertainties of the measurements than the FWHM does.
The gradual increase of W$_{eff}$, accompanied with the red-shift of the center of gravity of the profile, may suggest populating of higher transitions of P-branch of the bands of molecules, assuming the latter are DIB carriers. It is also shown, that diffuse bands are broader although more shallow in the harsh conditions of $\sigma$-clouds, where atomic and molecular lines are weakened and/or totally absent.  The difference of the effective width of DIBs in $\zeta$ and $\sigma$ clouds is discussed as well.
\end{abstract}

\maketitle

\section{Introduction}

The interstellar space, especially inside the disc of our Galaxy (and similar spiral galaxies), is filled with clouds of diffuse
matter. The average density of this environment is just one hydrogen atom per ccm - the so called "Oort limit". Optically
thicker regions (interstellar clouds) are usually divided into four types, according to their density: "diffuse", "translucent",
"dense", and star forming regions. Dense (and star forming) clouds can only be analysed in microwave or far-infrared spectral ranges;
in the optical/UV ones they are too opaque.
The presence of translucent clouds, the objects studied within the framework of this paper, is revealed mostly by optical,
absorption spectral features, present in spectra of distant stars, shining through the interstellar medium (ISM). The ISM consists of
several components: gas atoms, molecules and dust grains, being revealed by their absorption spectral features.

\begin{figure}
\centering
\includegraphics[width=8cm]{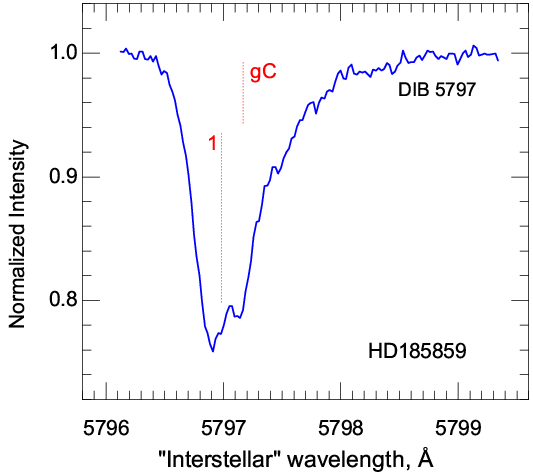} 
\caption{Asymmetric profile of DIB 5797 in a spectrum of HD185859. Note the different position of a conventional rest
wavelength position (1) adopted from Galazutdinov et al. (2000) and the center of gravity (gC) of the feature.}
\label{lambda0-vs-gC}
\end{figure}

\begin{figure*}
\centering
\includegraphics[width=14cm]{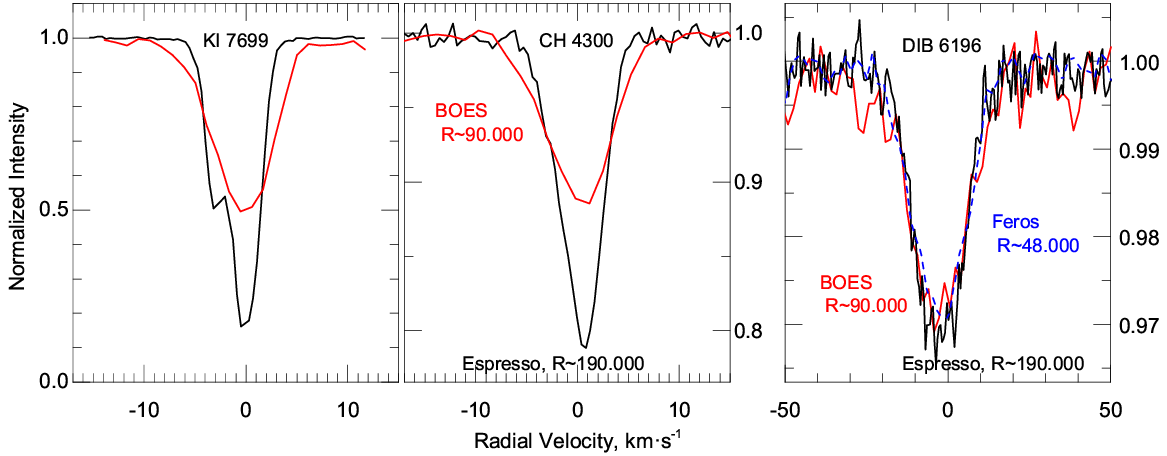} 
\caption{Comparison of profiles in spectra of HD23180 observed with different instruments and resolving power. Note the similarity of profiles
of the narrowest known diffuse band 6196 \AA\ observed with different resolving power (right panel). Despite of small differences in the profile of
 DIB 6196 there are no troubles with finding its center which is the same in all spectra.}
\label{resolvingpower}
\end{figure*}

The most common of the interstellar absorptive structures, probably seen in all diffuse and translucent clouds, is the Ca{\sc ii} doublet
(consisting of the so-called $H$ and $K$ lines). Its interstellar origin is known since the observations by \citep{Hartmann1904}.
The $H$ and $K$ lines are frequently observed in the absence of other interstellar features (perhaps, except the Na{\sc i}
ones). Many tiny, optically thin, diffuse clouds, seem to fill the galactic disc quite evenly \citep{2015PASP..127..126G}. Spectral
lines of almost all known elements can in fact be observed in the interstellar medium, but a vast majority of them is detectable only
due to resonance lines, situated in the vacuum ultraviolet, requiring space-born instruments to be observed.

Among the identified interstellar spectral lines and bands one can observe puzzling, pretty broad spectral features, known as diffuse
interstellar bands (DIBs) - known as unidentified since 1922 \citep{Heger1922}, believed to be carried by complex chemical species, including
prebiotic ones. In this paper we move one more step towards the characterization and identification of such features, focusing on the most
promising ones. DIB molecular origin is evidenced by the presence of substructures inside their profiles (\cite{Sarre1995},
\cite{Kerr1998}, \cite{2008PASP..120..178G}, \cite{2020AJ....159..113G}).
To identify their carriers it's necessary to compare astrophysical data with laboratory gas phase spectra of (for example) carbon
chains or Polycyclic Aromatic Hydrocarbons - PAHs (\cite{Sarre1995}, \cite{Salama1999}, \cite{motyl2000}). A vast majority of DIBs \citep{Fan 2019}
are narrow and shallow features; an analysis of their profile shapes requires a very high resolution and signal-to-noise (S/N) ratio.
The profiles are to be established in a purely observational way to avoid a bias if an attempt to find some
features, known from the lab or theoretical modeling, is put forward.

During the last three decades \cite{KG99}, \cite{G06}, \cite{2008PASP..120..178G} and \cite{2015MNRAS.451.3210K}
found red- and blue-shifts of some DIBs vs. atomic lines. Also their profiles (widths, shapes) apparently vary from one to another specific
environment. In some cases this likely follows variations of rotational temperatures of simplest carbon chains (C$_2$, C$_3$) -- see e.g.
Ka\'zmierczak et al. (2009). Experiments say that spectra of molecules (candidates to carriers of DIBs) should contain one strong and several weak
features each. It is observationally very challenging to relate weak DIBs to the strong ones and thus to divide them into
"families" sharing the same carrier (e.g. \cite{KW88}, \citep{KW87}). The S/N ratio inside the profiles of weak DIBs is
typically very low which makes their intensity measurements uncertain.

To measure reliably strengths of weak members of these "families" one needs a very high S/N ratio
and resolution (to separate the investigated DIBs from another stellar and interstellar features). This motivates the use of large
telescopes. The DIB contours can vary with temperature (kinetic and rotational) in the way characteristic to a given species. Changes
of profiles may lead to seemingly different central wavelengths. It is important to relate DIB profiles' widths, substructure patterns
and intensities to abundances of CH, CH$^+$, CN and rotational
temperatures of the C$_2$ and C$_3$ molecular species.  They are likely related if the DIBs are carried by some molecules.

Until now the wavelength variability of diffuse bands were observed in three different cases:
\begin{itemize}
\item blue-shift of diffuse bands. It was reported for HD34078 and objects in Sco OB1 association (\citet{2015PASP..127..126G}; \citet{2008PASP..120..178G}; \citet{G06}).
\item red-shift of diffuse bands \citep{2015MNRAS.451.3210K}. The wavelength displacements of the diffuse interstellar bands at 4502, 5705, 5780, 6284 and 7224 \AA,
  with respect to the well-known, narrow atomic/molecular interstellar lines, have been measured in the spectra of the two Orion Trapezium stars.
\item broadening of profiles of diffuse bands. In  \cite{2021MNRAS.508.4241K} we reported the profile broadening and  peak wavelength variation for 5780, 5797, 6196 and 6614 \AA\ DIBs
and found strong variability sometimes doubling the features' width. The profile broadening in the studied DIBs moves the profile's
centers towards longer wavelengths, probably due to the excitation of higher levels of the P branch of the unknown molecular carrier's band.
Moreover, diffuse bands are broader in clouds with abundantly populated vibrationally excited states of the hydrogen molecules, i.e. DIB's broadening correlates with the
rotational temperature estimated on H$_2$ $\nu$=2 vibrational level.
\end{itemize}

Here we returned to the major DIB 5780 and 5797 
\footnote{Major DIBs serve to divide interstellar clouds to $\sigma$ and $\zeta$ types.
 Lines of sight, referred to as $\sigma$ clouds (because the line of sight toward $\sigma$ Scorpii is the archetype), are characterized by a low strength ratio of DIBs 5797/5780. Conversely, in $\zeta$-clouds (after $\zeta$ Ophiuchi), one can expect a high ratio of these DIBs \citep{KW88}. Moreover, $\zeta$-type clouds show usually strong absorption lines of interstellar molecules such as CH, CH$^+$, CN, C$_2$, C$_3$, etc., whereas  the latter are weak in $\sigma$-type clouds (in many cases - below the level of detection).}
using the "interstellar"\, wavelength scale based on the narrowest known DIB 6196 and,
we provide a similar study for DIBs 6284 and 7224~\AA. Also we classified diffuse bands using the effective width - a parameter, originally introduced by us in 2008
\citep{2008PASP..120..178G}. Results of the new approach applied to study the variability of the wavelengths of diffuse bands are given below.

\section{Observational material}

The observational data (Table 1) have been collected using several high resolution, echelle spectrographs:
\begin{itemize}
\item
     UVES (Ultraviolet and Visual Echelle Spectrograph) fed by the 8m Kueyen VLT mirror
     \citep{Dekal00}. The spectral resolution is up to R $\equiv$ $\lambda/\Delta\lambda$  = 80,000 in the blue range and R = 110,000 in the red one.
     The telescope size allows to get high S/N ratio spectra of even pretty faint stars. The selected spectra have been collected
     in the frame of the Large Program EDIBLES (ESO Diffuse Interstellar Bands Large Exploration Survey) and our previous study
     \citep{Siebenmorgen2020}  where we provide online access to the analyzed UVES data.\footnote{https://vizier.u-strasbg.fr/viz-bin/VizieR?-source=J/A+A/641/A35}
\item
     Some southern objects were studied with the aid of Feros - fiber fed echelle
     spectrograph \citep{Ketal99} at the ESO's La Silla observatory in Chile. Feros
     provides the resolving power R$\equiv$$\lambda$/$\Delta$$\lambda$
     of 48,000 and allows to get the whole available spectral range
     ($\sim$3700 -- 9200~\AA, divided into 37 orders) recorded in a
     single exposure.
\item
    ESPaDOnS spectrograph (Echelle SpectroPolarimetric Device for the
    Observation of Stars)\footnote{https://www.cfht.hawaii.edu/Instruments/Spe\-ctroscopy/Espadons/} is the bench-mounted high-resolution echelle
    spectrograph/spec\-tro\-pola\-ri\-meter) attached to the 3.58~m
    Canada-France-Hawaii telescope (CFHT) at Mauna Kea (Hawaii, USA). It is
    designed to obtain a complete optical spectrum in the range from
    3,700 to 10,050~~\AA. The whole spectrum is divided into 40 \'{e}chelle
    orders. The resolving power is about 68,000.
\item
    MIKE spectrograph  \citep{Ber03} fed by the Magellan/Clay telescope of the Las Campanas Observatory in Chile.
    The spectral resolution with a 0.35$\times$5 arcs slit ranges from
    $\sim$56,000 on the blue side (3600-5000~~\AA) to $\sim$77,000 on the red side (4800-9400~~\AA).
\item
    The Bohyunsan Echelle Spectrograph (BOES) of the Korean National Observatory \citep{kimetal2007} is installed at the 1.8m telescope of the Bohyunsan Observatory in Korea. The spectrograph has three
    observational modes allowing resolving powers of 30,000, 45,000, and
    90,000. In any mode, the spectrograph covers the whole spectral range of
    $\sim$3500 to $\sim$10,000~\AA, divided into 75 -- 76 spectral orders.
\item
    ESPRESSO (Echelle Spectrograph for Rocky Exoplanets and Stable Spectroscopic Observations) \citep{PMC14} is a highly-stabilized fibre-fed echelle spectrograph that can be fed with light from either a single or
    up to four Unit Telescopes (UTs) simultaneously.\footnote{https://www.eso.org/sci/facilities/paranal/instruments/espresso.html}
    The instrument is installed at the incoherent combined Coude focus (ICCF) of the VLT.
    The spectrograph is fed by two fibres, one for the scientific target and the other one for simultaneous reference (either the sky or a
    simultaneous drift reference, the Fabry-Perot). The light from the two fibres is recorded onto a blue (380-525nm) and a red (525-788nm)
    CCD mosaic. The spectral resolution is either 190,000, 140,000 or 70,000 and does not influence the spectral range covered.
\end{itemize}

We processed raw data and made measurements in the reduced spectra with our interactive analysis software DECH \citep{dech}.
  For the DECH data reduction, we averaged bias images for subsequent correction of  all other images. The scattered light was determined as a complex
  shaped two-dimensional surface function, which is individually   calculated for each stellar and flat-field frame by a two-dimensional cubic-spline
  approximation over areas of minima between the spectral orders.  Then,   the pixel-to-pixel variations across the CCD were corrected by
  dividing all stellar frames by the averaged and normalized flat-field   frame. One-dimensional stellar spectra were extracted by simple
  summation in the cross-dispersion direction along the width of each   spectral order. The extracted spectra of the same object, observed in
  the same night, were averaged to achieve the highest signal-to-noise ratio.
  Fiducial continuum normalization was based on a cubic spline  interpolation over the interactively selected anchor points.

The wavelength scale was constructed on the basis of a global polynomial of the form:
$$
\lambda(x,m)=\sum_{i=0}^{k}\sum_{j=0}^{n} a_{ij} x^{i} m^{j}
$$
where $a_{ij}$ are polynomial coefficients; $x$ - the pixel position in the direction of dispersion;  $m$ - the order
number. The final solution uses typically 1000 - 1500 ThAr lines and the rms residual error between the fit and
the position of the lines is usually $\leq$0.003~\AA. Then, the wavelength scale was converted to the rest wavelength interstellar frame using the narrowest
known diffuse interstellar band at 6195.97 \AA\ \citep{2000PASP..112..648}.

 Along with equivalent width, FWHM, etc., DECH software provides estimates of the center of gravity (gC) of absorption line profiles in continuum-normalized spectra S$_{\lambda}$:
$$
  gC = \sum_{\lambda=\lambda_1}^{\lambda_2} (1-S_{\lambda})\lambda / \sum_{\lambda=\lambda_1}^{\lambda_2} (1-S_{\lambda})
$$

Fig.1 exhibits both conventional rest wavelength of diffuse bands 5797 \AA\ (adopted from Galazutdinov et al. (2000)) and the center of gravity of this evidently asymmetric feature.

The analysed spectra are originated from instruments with different resolving power, sometimes not sufficient for studies of profiles of very narrow
atomic/molecular interstellar lines and details of profiles of diffuse bands though being good enough for measurement of the center of gravity, intensity (depth)
and equivalent width of diffuse bands (Fig. 2).

\begin{figure*} 
\centering
\includegraphics[width=14cm]{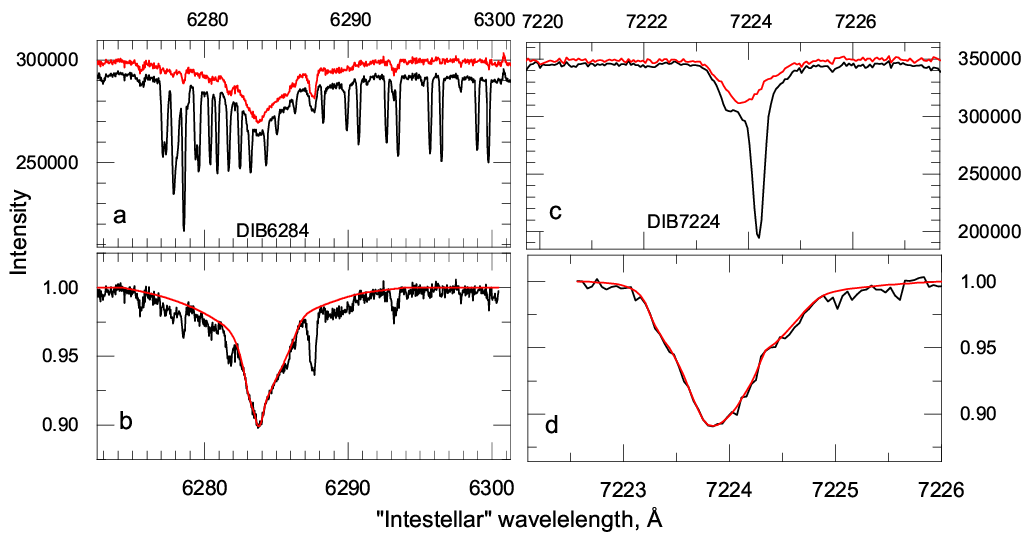}
\caption{Profiles of diffuse bands 6284 \AA\ (a,b) and 7224 \AA\ (c, d) in spectra of HD185859 observed with fiber-fed BOES spectrograph.
{\bf a)} Original spectrum in the area of very broad DIB 6284 (black) and, the same spectrum after removal of telluric lines (red);
{\bf b)} Example of measurement of complex shaped DIB 6284 blended with other features of stellar and interstellar origin (black)
with an interactively constructed fiducial profile (red smooth line);
{\bf c)} Original spectrum in the area of DIB 7224 (black) and, the same spectrum after removal of telluric lines (red);
{\bf d)} Cleaned DIB 7224 profile (black) with an interactively constructed fiducial profile (red smooth line);}
\label{hd185859}
\end{figure*}

\begin{figure*} 
\centering
\includegraphics[width=14cm]{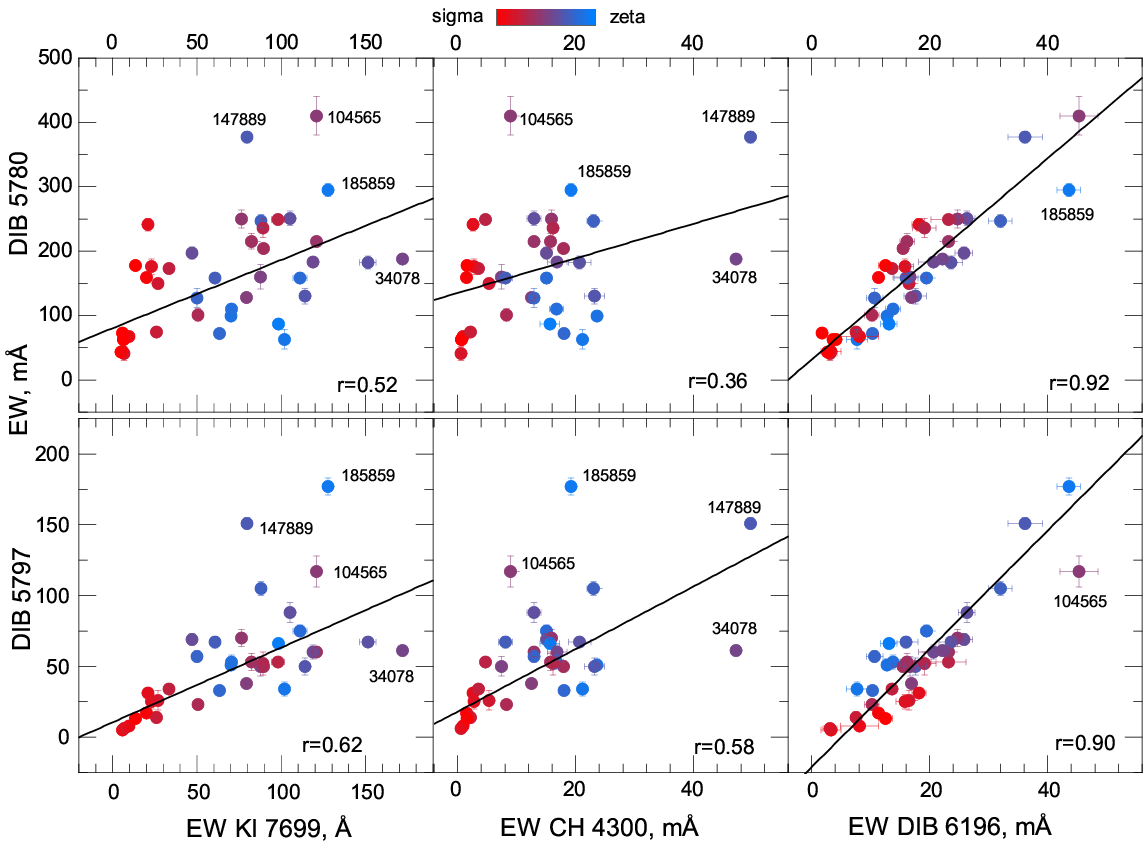}
\caption{Correlation between equivalent width of major DIBs 5797 and 5780 with interstellar CH 4300 \AA, K{\sc i} and DIB 6196. Pearson correlation coefficient is given in each plot.
 The colors interpreted at the top of the plot show the ratio of equivalent width of major DIBs 5797/5780, i.e. the blue circles are $\zeta$-objects while the red ones are $\sigma$ ones.
}
\label{EWmajorsKICH6196}
\end{figure*}

\begin{figure*} 
\centering
\includegraphics[width=14cm]{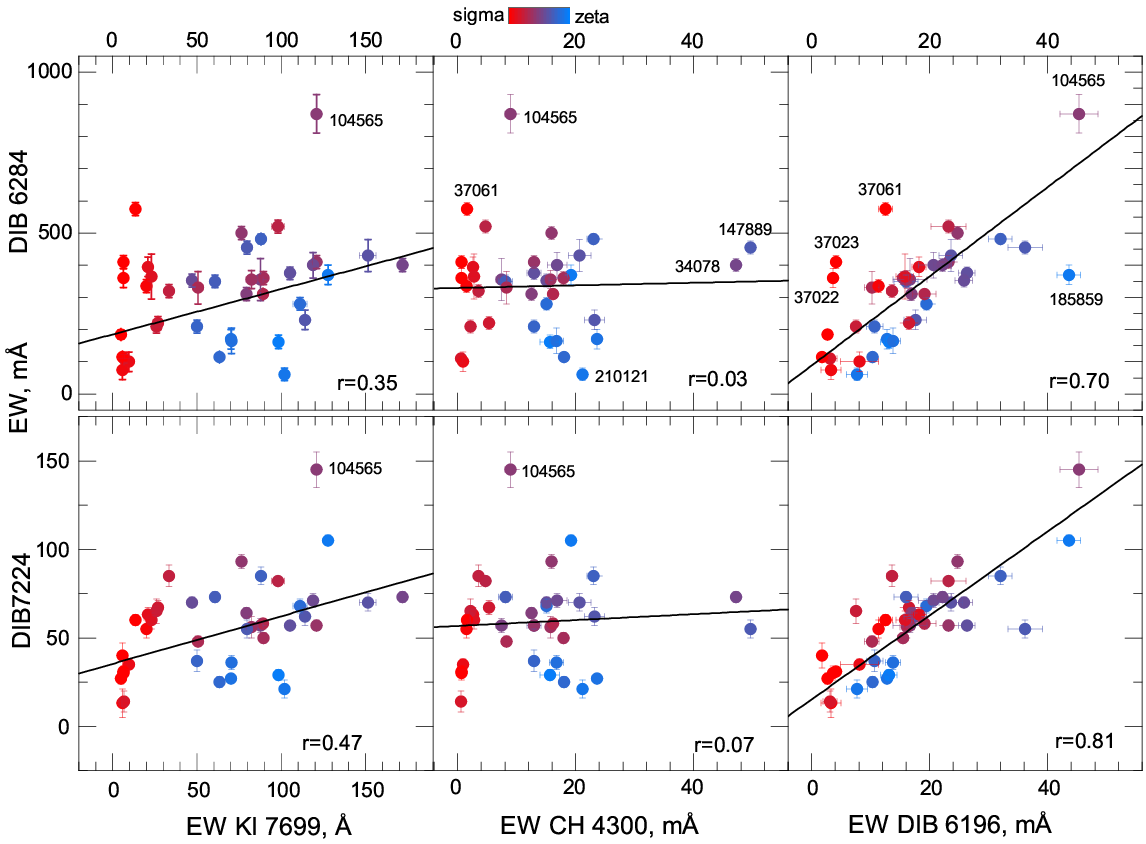}
\caption{The same as in Fig. 4 for diffuse bands 6284 and 7224.}
\label{68247224KICH6196}
\end{figure*}

\begin{figure} 
\centering
\includegraphics[width=8cm]{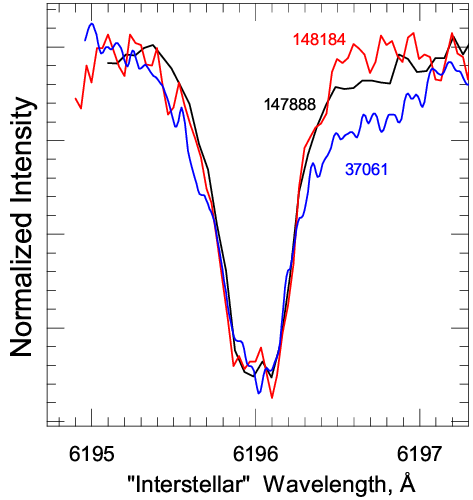}
\caption{Profile of DIB 6196 with different effective width W$_{eff}$. Profiles are normalized to the same depth for clarity.
Note that variability mostly occurs in the red wing of the profile while the width of profile in the lower part is hardly variable
providing unambiguous estimate of its center.}
\label{prf6196}
\end{figure}

\begin{figure} 
\centering
\includegraphics[width=8cm]{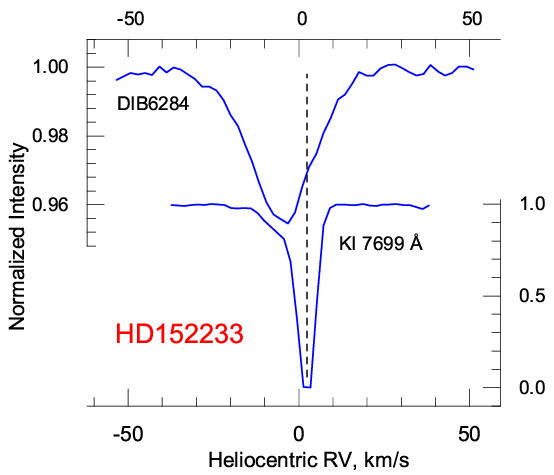}
\caption{The difference of positions of interstellar features in spectrum of a Sco OB1 object HD152233.}
\label{hd152233}
\end{figure}

\begin{figure*} 
\centering
\includegraphics[width=14cm]{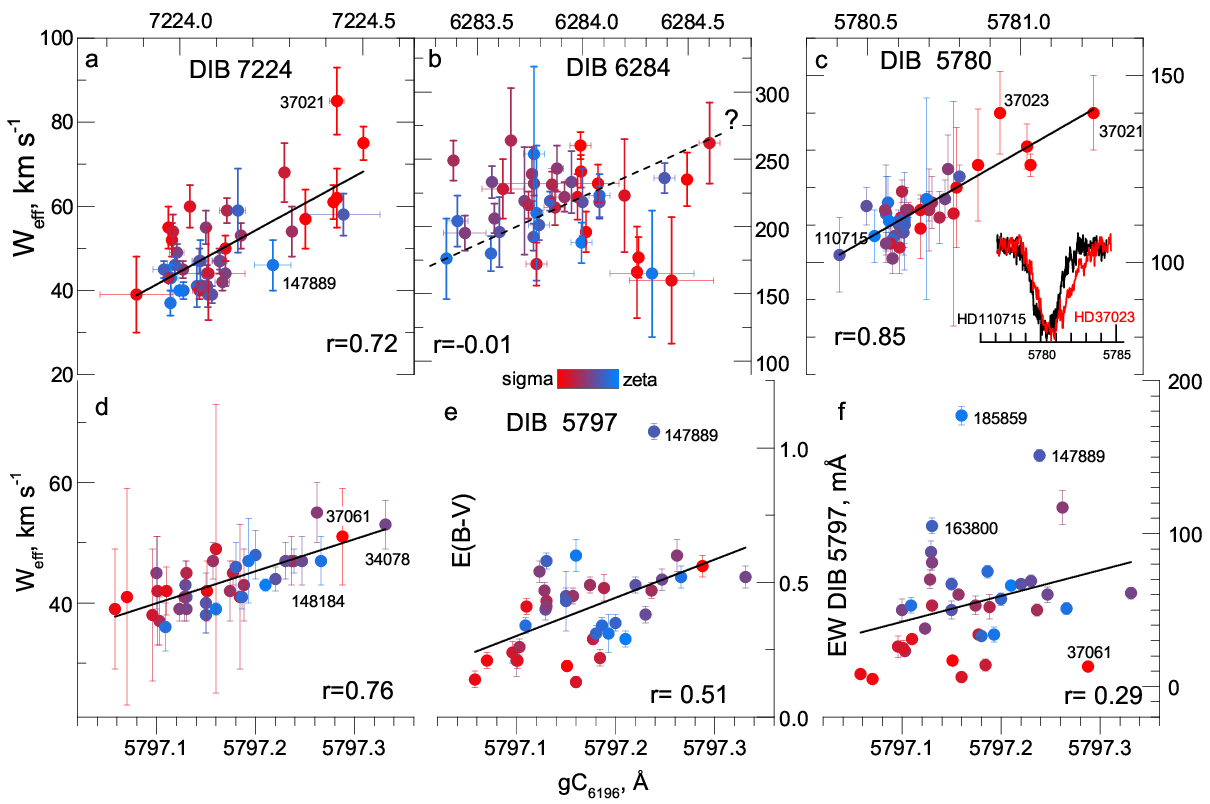}
\caption{{\bf (a, b, c, d)} Relations between the effective width (W$_{eff}$) and the center of gravity (gC) of diffuse bands. The "interstellar"\, wavelength scale based on DIB 6195.97 \AA.
DIB 5780 profiles shown for two objects (in panel {\bf (c)}) exhibits both real red-shift and broadening; {\bf (e)} The center of gravity of DIB 5797 versus E$_{B-V}$;
{\bf (f)} The center of gravity of DIB 5797 versus its equivalent widths. The color shows the EW ratio of major DIBs (5797/5780). The blue symbols are $\zeta$-clouds while the red ones are $\sigma$-clouds.}
\label{GCWeff}
\end{figure*}

Both DIBs 6284 and 7224 are severely polluted by telluric lines. In order of cleaning DIB profiles from telluric contamination we observed so called "divisors"\,  - spectra of hot rapidly
rotating stars free of interstellar reddening. In Fig. 3 we provide an example of the telluric lines removal technic applied for spectra of HD185859, where original (polluted) spectrum
was divided by spectrum of Spica. DIB 6284 is one of the broadest and strongest known diffuse bands, always blended with other interstellar and/or telluric lines.
We measured the equivalent width of these complex shaped profiles (sometime with remnants of telluric lines) using an interactively constructed fiducial profile (Fig. 3).
Intensity (depth) of the latter was used to calculate the effective width W$_{eff}$ which is the ratio of the equivalent width of a line to its central depth. The procedure is implemented in the DECH software \citep{dech}.

Profile of DIB 7224 is also polluted by telluric lines although is much more narrower than that of the DIB 6284. Nevertheless we used the same method for measurements of W$_{eff}$ and equivalent width.
Owing to the smoothness of the fiducial profile the impact of the noise pattern on the accuracy of measurements was reduced.

\begin{sidewaystable}
\centering
\caption{Basic parameters of selected stars and measurements data: equivalent widths,  the effective wavelength  $\lambda_{gc}$ (the center of gravity of the line profile), the effective width  W$_{eff}$ of DIBs,
 the equivalent widths (m\AA). Ending letters in the star names mark the origin of the analyzed spectra: b - BOES, e - ESPADONS, s- Espresso, f- FEROS, m - MIKE, u - UVES. \\
}
\resizebox{\textwidth}{!}{
\begin{tabular}{llcccccccccccccccc}
\hline
HD    & Sp/L  &  V    & E$_{B-V}$        & EW$_{6284}$&$\lambda_{gc}6284$
                                                                        &W$_{eff}6284$&EW$_{7224}$
                                                                                                &$\lambda_{gc}7224$
                                                                                                                 &W$_{eff}7224$&
                                                                                                                           EW$_{4300}$  &EW$_{7699}$  &EW$_{5780}$ &$\lambda_{gc}5780$
                                                                                                                                                                            &W$_{eff}$5780
                                                                                                                                                                                        & EW$_{5797}$
                                                                                                                                                                                                    &$\lambda_{gc}5797$
                                                                                                                                                                                                            &W$_{eff}$5797\\
         &        & [mag] & [mag]         &  [m\AA]    &   [\AA]        &[\kms]      & [m\AA]   & [\AA]          &[\kms]   &[m\AA]      &  [m\AA]     &  [m\AA]    &[\AA]   &[\kms]     &  [m\AA]   & [\AA] &[\kms]\\
\hline
22951u  &   B0.5V & 4.97 & 0.26$\pm$0.03 &330$\pm$50 & 6283.66$\pm$0.01 & 264$\pm$39 & 48$\pm$3 &7224.11$\pm$0.02&47$\pm$3 & 8.3$\pm$0.8& 50.6$\pm$3.0& 101$\pm$10 &5780.70 & 114$\pm$11&  23$\pm$3 &5797.10& 37$\pm$4  \\
23180b  &   B1III & 3.86 & 0.29$\pm$0.03 &162$\pm$21 & 6283.77$\pm$0.05 & 254$\pm$65 & 29$\pm$3 &7223.98$\pm$0.02&43$\pm$3 &15.6$\pm$1.6& 98.5$\pm$2.5&  87$\pm$5  &5780.52 & 107$\pm$7 &  66$\pm$2 &5797.21& 43$\pm$1  \\
24263u  &   B3.5V & 5.77 & 0.21$\pm$0.06 &355$\pm$65 & 6283.72$\pm$0.02 & 219$\pm$40 & 57$\pm$4 &7224.12$\pm$0.01&44$\pm$3 & 7.4$\pm$1.0& 87.6$\pm$1.5& 160$\pm$19 &5780.56 & 113$\pm$13&  50$\pm$7 &5797.10& 45$\pm$6  \\
24398bh &   B1Ib  & 2.85 & 0.34$\pm$0.03 &165$\pm$40 & 6283.78$\pm$0.03 & 210$\pm$50 & 36$\pm$4 &7224.06$\pm$0.02&47$\pm$5 &16.8$\pm$1.2& 70.6$\pm$2.7& 110$\pm$6  &5780.57 & 116$\pm$7 &  53$\pm$5 &5797.11& 36$\pm$4  \\
30492u  &   B9.5V & 9.02 & 0.40$\pm$0.04 &375$\pm$20 & 6283.77$\pm$0.01 & 192$\pm$10 & 57$\pm$2 &7224.09$\pm$0.01&39$\pm$2 &13.0$\pm$1.2&105.3$\pm$2.5& 251$\pm$11 &5780.50 & 115$\pm$5 &  88$\pm$7 &5797.13& 43$\pm$4  \\
34078m  &   O9.5V & 5.96 & 0.52$\pm$0.04 &400$\pm$20 & 6283.57$\pm$0.02 & 233$\pm$12 & 73$\pm$3 &7223.99$\pm$0.02&49$\pm$2 &47.2$\pm$0.6&171.8$\pm$1.9& 188$\pm$7  &5780.75 & 117$\pm$5 &  61$\pm$4 &5797.33& 53$\pm$4  \\
37020me &   B0V   & 6.73 & 0.30$\pm$0.03 &185$\pm$14 & 6284.01$\pm$0.02 & 196$\pm$15 & 27$\pm$3 &7224.43$\pm$0.01&62$\pm$7 & $<$0.3     &  4.9$\pm$1.3&  43$\pm$5  &5780.86 & 126$\pm$15&           &       &           \\
37021e  &   B1V   & 7.96 & 0.50$\pm$0.05 &115$\pm$10 & 6284.26$\pm$0.01 & 177$\pm$15 & 40$\pm$7 &7224.43$\pm$0.02&85$\pm$8 & $<$0.6     &  5.9$\pm$0.4&  73$\pm$5  &5781.24 & 140$\pm$10&           &       &           \\
37022emh&   O8V   & 5.13 & 0.34$\pm$0.04 &360$\pm$30 & 6284.50$\pm$0.01 & 235$\pm$20 & 30$\pm$4 &7224.34$\pm$0.01&57$\pm$7 & 0.7$\pm$0.1&  6.4$\pm$0.8&  63$\pm$3  &5781.02 & 131$\pm$6 &           &       &           \\
37023e  &   B1.5V & 6.70 & 0.34$\pm$0.04 &410$\pm$20 & 6283.99$\pm$0.02 & 241$\pm$12 & 31$\pm$2 &7224.42$\pm$0.01&61$\pm$4 & 0.7$\pm$0.2&  6.3$\pm$0.8&  63$\pm$5  &5780.93 & 140$\pm$11&           &       &           \\
37041u  &   O8V   & 6.30 & 0.21$\pm$0.03 & 74$\pm$30 & 6284.26$\pm$0.02 & 166$\pm$34 & 13$\pm$8 &7223.88$\pm$0.1 &39$\pm$9 & $<$0.4     &  5.8$\pm$1.0&  44$\pm$6  &5780.79 & 120$\pm$16&   5$\pm$3 &5797.07& 41$\pm$18 \\
37061e  &   O9V   & 6.83 & 0.56$\pm$0.04 &575$\pm$20 & 6283.99$\pm$0.05 & 260$\pm$10 & 60$\pm$3 &7224.50$\pm$0.01&75$\pm$4 & 1.6$\pm$0.2& 13.5$\pm$0.5& 178$\pm$5  &5781.03 & 126$\pm$3 &  13$\pm$2 &5797.29& 51$\pm$8  \\
104565u &  O9.7II & 9.25 & 0.60$\pm$0.06 &870$\pm$60 & 6283.88$\pm$0.03 & 243$\pm$17 &145$\pm$10&7224.07$\pm$0.01&55$\pm$4 & 9.0$\pm$1.0&121.0$\pm$2.2& 410$\pm$30 &5780.76 & 125$\pm$9 & 117$\pm$11&5797.26& 55$\pm$5  \\
110432u &  B0.5IV & 5.31 & 0.54$\pm$0.04 &310$\pm$20 & 6283.44$\pm$0.10 & 195$\pm$13 & 64$\pm$3 &7224.07$\pm$0.01&44$\pm$2 &12.6$\pm$0.4& 79.2$\pm$1.6& 128$\pm$5  &5780.58 & 101$\pm$4 &  38$\pm$2 &5797.12& 39$\pm$2  \\
110715u &  B9V    & 8.65 & 0.45$\pm$0.03 &230$\pm$30 & 6283.60$\pm$0.01 & 196$\pm$26 & 62$\pm$5 &7224.08$\pm$0.01&39$\pm$3 &23.2$\pm$1.7&114.1$\pm$1.6& 130$\pm$12 &5780.41 & 102$\pm$10&  50$\pm$6 &5797.15& 40$\pm$5  \\
133518u &  B2IV   & 6.39 & 0.14$\pm$0.03 &100$\pm$30 & 6284.42$\pm$0.20 & 160$\pm$47 & 35$\pm$3 &7223.98$\pm$0.01&52$\pm$4 & 0.9$\pm$0.4&  9.6$\pm$0.6&  67$\pm$5  &5780.67 & 109$\pm$8 &   8$\pm$2 &5797.06& 39$\pm$10 \\
143275ub&  B0.3IV & 2.32 & 0.22$\pm$0.03 &210$\pm$20 & 6283.62$\pm$0.10 & 228$\pm$22 & 65$\pm$7 &7224.29$\pm$0.01&68$\pm$7 & 2.2$\pm$0.7& 25.9$\pm$0.7&  74$\pm$5  &5780.60 & 104$\pm$7 &  14$\pm$4 &5797.18& 41$\pm$12 \\
144217ub&  B1V    & 2.62 & 0.19$\pm$0.02 &335$\pm$20 & 6284.07$\pm$0.10 & 232$\pm$14 & 55$\pm$5 &7223.97$\pm$0.01&55$\pm$5 & 1.5$\pm$0.7& 20.0$\pm$2.0& 159$\pm$5  &5780.62 & 110$\pm$3 &  17$\pm$2 &5797.15& 42$\pm$5  \\
144470ub&  B1V    & 3.97 & 0.21$\pm$0.03 &365$\pm$70 & 6284.20$\pm$0.01 & 223$\pm$42 & 60$\pm$5 &7224.03$\pm$0.01&60$\pm$5 & 2.8$\pm$0.7& 23.1$\pm$2.0& 176$\pm$12 &5780.61 & 112$\pm$8 &  25$\pm$5 &5797.10& 42$\pm$9  \\
145502m &  B2V    & 4.00 & 0.29$\pm$0.02 &320$\pm$20 & 6283.87$\pm$0.05 & 214$\pm$13 & 85$\pm$6 &7223.98$\pm$0.01&54$\pm$4 & 3.6$\pm$0.1& 33.5$\pm$1.3& 173$\pm$3  &5780.63 & 114$\pm$2 &  34$\pm$2 &5797.18& 45$\pm$2  \\
147165u &  B1III  & 2.89 & 0.41$\pm$0.03 &395$\pm$30 & 6283.97$\pm$0.01 & 222$\pm$17 & 63$\pm$4 &7224.12$\pm$0.01&50$\pm$3 & 2.6$\pm$0.5& 20.9$\pm$1.5& 241$\pm$10 &5780.67 & 114$\pm$4 &  31$\pm$3 &5797.11& 42$\pm$4  \\
147888uf&  B3V    & 6.74 & 0.48$\pm$0.05 &310$\pm$10 & 6283.85$\pm$0.02 & 231$\pm$10 & 58$\pm$3 &7224.13$\pm$0.02&59$\pm$3 &16.1$\pm$0.3& 89.3$\pm$1.1& 236$\pm$15 &5780.73 & 112$\pm$7 &  52$\pm$8 &5797.19& 43$\pm$6  \\
147889uf&  B2.5V  & 7.90 & 1.06$\pm$0.03 &455$\pm$20 & 6284.39$\pm$0.05 & 236$\pm$11 & 55$\pm$5 &7224.45$\pm$0.1 &58$\pm$5 &49.6$\pm$0.6& 79.7$\pm$0.9& 377$\pm$9  &5780.80 & 123$\pm$3 & 151$\pm$4 &5797.24& 47$\pm$2  \\
147932f &  B5V    & 7.27 & 0.49$\pm$0.03 &355$\pm$28 & 6283.76$\pm$0.02 & 239$\pm$20 & 56$\pm$6 &7224.31$\pm$0.01&54$\pm$6 &15.7$\pm$0.8& 82.4$\pm$3.4& 215$\pm$12 &5780.71 & 116$\pm$7 &  53$\pm$6 &5797.17& 42$\pm$5  \\
147933e &  B1V    & 5.05 & 0.47$\pm$0.03 &360$\pm$20 & 6283.38$\pm$0.02 & 249$\pm$15 & 50$\pm$3 &7224.17$\pm$0.01&53$\pm$3 &18.0$\pm$0.6& 89.5$\pm$0.9& 204$\pm$7  &5780.70 & 115$\pm$4 &  50$\pm$4 &5797.24& 47$\pm$4  \\
148184ue&  B2V    & 4.43 & 0.52$\pm$0.04 &170$\pm$30 & 6283.35$\pm$0.02 & 176$\pm$30 & 27$\pm$2 &7223.97$\pm$0.01&37$\pm$3 &23.6$\pm$0.8& 70.1$\pm$0.9&  99$\pm$7  &5780.62 & 112$\pm$8 &  51$\pm$4 &5797.27& 47$\pm$4  \\
\hline
\end{tabular}
}
\label{targets}
\end{sidewaystable}

\setcounter{table}{0}
\begin{sidewaystable}
\centering
\caption{continued.
}
\resizebox{\textwidth}{!}{
\begin{tabular}{llcccccccccccccccc}
\hline
HD    & Sp/L  &  V    & E$_{B-V}$        & EW$_{6284}$&$\lambda_{gc}6284$
                                                                        &W$_{eff}6284$&EW$_{7224}$
                                                                                                &$\lambda_{gc}7224$
                                                                                                                 &W$_{eff}7224$&
                                                                                                                           EW$_{4300}$  &EW$_{7699}$  &EW$_{5780}$ &$\lambda_{gc}5780$
                                                                                                                                                                            &W$_{eff}$5780
                                                                                                                                                                                        & EW$_{5797}$
                                                                                                                                                                                                    &$\lambda_{gc}5797$
                                                                                                                                                                                                            &W$_{eff}$5797\\
         &        & [mag] & [mag]         &  [m\AA]    &   [\AA]        &[\kms]      & [m\AA]   & [\AA]          &[\kms]   &[m\AA]      &  [m\AA]     &  [m\AA]    &[\AA]   &[\kms]     &  [m\AA]   & [\AA] &[\kms]\\
\hline
148579u &  B8V    & 7.32 & 0.35$\pm$0.03 &210$\pm$20 & 6283.40$\pm$0.05 & 204$\pm$19 & 37$\pm$6 &7224.16$\pm$0.02&59$\pm$10&12.9$\pm$0.8& 50.0$\pm$0.9& 127$\pm$15 &5780.62 & 111$\pm$13&  57$\pm$4 &5797.20& 48$\pm$4  \\
148605uf&  B3V    & 4.79 & 0.13$\pm$0.02 &110$\pm$14 & 6284.60$\pm$0.05 & 262$\pm$30 & 14$\pm$6 &7224.08$\pm$0.1 &44$\pm$11& 0.6$\pm$0.3&  6.5$\pm$0.5&  41$\pm$11 &5780.78 & 113$\pm$30&   6$\pm$3 &5797.16& 49$\pm$24 \\
149757e &  O9.2IV & 2.56 & 0.31$\pm$0.03 &114$\pm$8  & 6283.56$\pm$0.01 & 180$\pm$13 & 25$\pm$3 &7224.05$\pm$0.01&41$\pm$5 &18.0$\pm$0.3& 63.2$\pm$0.7&  72$\pm$6  &5780.56 & 113$\pm$10&  33$\pm$3 &5797.18& 46$\pm$4  \\
152076f &  B0/1III& 8.90 & 0.49$\pm$0.03 &430$\pm$50 & 6284.00$\pm$0.02 & 218$\pm$25 & 70$\pm$5 &7224.05$\pm$0.02&47$\pm$3 &20.7$\pm$1.9&151.3$\pm$4.9& 182$\pm$10 &5780.62 & 108$\pm$6 &  67$\pm$3 &5797.22& 44$\pm$2  \\
152218uf&  O9V    & 7.57 & 0.51$\pm$0.04 &400$\pm$40 & 6283.95$\pm$0.03 & 233$\pm$23 & 71$\pm$4 &7224.11$\pm$0.02&47$\pm$2 &16.9$\pm$1.7&118.9$\pm$1.7& 183$\pm$5  &5780.60 & 110$\pm$3 &  60$\pm$5 &5797.25& 47$\pm$4  \\
152233mf&  O6II(f)& 6.59 & 0.45$\pm$0.03 &410$\pm$20 & 6283.91$\pm$0.04 & 222$\pm$11 & 57$\pm$3 &7224.01$\pm$0.01&45$\pm$2 &12.9$\pm$0.6&120.8$\pm$2.4& 215$\pm$3  &5780.56 & 114$\pm$2 &  60$\pm$3 &5797.16& 47$\pm$3  \\
152249uf&  O9.5Iab& 6.45 & 0.43$\pm$0.03 &520$\pm$20 & 6283.74$\pm$0.02 & 216$\pm$10 & 82$\pm$2 &7223.97$\pm$0.01&43$\pm$1 & 4.7$\pm$0.3& 98.1$\pm$3.9& 249$\pm$6  &5780.61 & 119$\pm$3 &  53$\pm$3 &5797.13& 45$\pm$2  \\
154445u &  B1IV   & 5.61 & 0.38$\pm$0.03 &352$\pm$20 & 6284.08$\pm$0.02 & 218$\pm$12 & 70$\pm$3 &7224.07$\pm$0.01&41$\pm$2 &15.1$\pm$0.4& 47.0$\pm$0.9& 197$\pm$9  &5780.59 & 107$\pm$5 &  69$\pm$4 &5797.23& 47$\pm$3  \\
163800u &  O7.5III& 7.00 & 0.58$\pm$0.03 &482$\pm$15 & 6283.84$\pm$0.02 & 219$\pm$7  & 85$\pm$5 &7223.96$\pm$0.03&45$\pm$2 &23.0$\pm$1.1& 87.8$\pm$1.5& 247$\pm$10 &5780.62 & 110$\pm$5 & 105$\pm$5 &5797.13& 41$\pm$2  \\
164906u &  B1IV   & 7.38 & 0.43$\pm$0.11 &350$\pm$20 & 6283.79$\pm$0.05 & 201$\pm$11 & 73$\pm$3 &7223.99$\pm$0.02&46$\pm$2 & 8.1$\pm$1.2& 60.8$\pm$2.4& 158$\pm$5  &5780.60 & 111$\pm$3 &  67$\pm$4 &5797.15& 38$\pm$3  \\
170740u &  B2V    & 5.72 & 0.47$\pm$0.03 &500$\pm$20 & 6283.77$\pm$0.20 & 232$\pm$9  & 93$\pm$4 &7224.12$\pm$0.02&42$\pm$2 &15.9$\pm$0.4& 76.3$\pm$1.1& 250$\pm$14 &5780.63 & 114$\pm$6 &  70$\pm$6 &5797.13& 41$\pm$4  \\
179406us&  B2.5II & 5.34 & 0.34$\pm$0.03 &280$\pm$20 & 6284.08$\pm$0.05 & 223$\pm$16 & 68$\pm$4 &7224.01$\pm$0.01&40$\pm$2 &15.1$\pm$0.8&111.1$\pm$4.2& 158$\pm$8  &5780.58 & 106$\pm$6 &  75$\pm$4 &5797.19& 41$\pm$2  \\
184915ub&  B0.5III& 4.96 & 0.24$\pm$0.04 &220$\pm$20 & 6283.78$\pm$0.05 & 172$\pm$16 & 67$\pm$4 &7224.05$\pm$0.02&40$\pm$2 & 5.3$\pm$0.4& 26.8$\pm$0.7& 150$\pm$7  &5780.57 & 105$\pm$5 &  26$\pm$7 &5797.10& 38$\pm$11 \\
185859ub&  B0Iab  & 6.59 & 0.60$\pm$0.06 &370$\pm$30 & 6283.99$\pm$0.05 & 188$\pm$15 &105$\pm$3 &7224.00$\pm$0.02&40$\pm$1 &19.2$\pm$0.6&127.7$\pm$2.6& 295$\pm$10 &5780.57 & 111$\pm$3 & 177$\pm$6 &5797.16& 39$\pm$1  \\
210121  &  B3.5V  & 7.68 & 0.31$\pm$0.07 & 61$\pm$20 & 6284.33$\pm$0.20 & 165$\pm$47 & 21$\pm$5 &7224.25$\pm$0.05&46$\pm$6 &21.1$\pm$0.6&102.0$\pm$2.0&  63$\pm$15 &5780.69 & 117$\pm$27&  34$\pm$5 &5797.19& 47$\pm$7  \\
\hline
\end{tabular}
}
\label{targets2}
\end{sidewaystable}

Let us summarize the base of new approaches we offer in the analysis of diffuse interstellar bands, namely in conversion of the wavelength scale into the interstellar frame,
in the estimate of the rest wavelength of diffuse bands and, in estimate of the widths of diffuse bands:
\begin{itemize}
  \item Equivalent widths of DIB 6196 and other diffuse bands demonstrate very good mutual correlations which is better than that existing one between EW's of diffuse bands and
        interstellar lines of K{\sc i} 7769 \AA\ or CH 4300 \AA\, -- the latter are usually serving as a reference for making the interstellar wavelength scale.
        The difference of the correlation magnitudes means that diffuse bands' carriers are occupying almost the same volumes of interstellar clouds while atomic gas K{\sc i}
        may be located apart (at least partially) of these volumes, i.e. DIB 6196 is a best reference point although the best does not mean perfect --
        carriers of diffuse bands are not mandatorily occupying the same volumes;
  \item Normally the position of a diffuse band is estimated using somewhat voluntary selected part of its profiles, generally being somewhere close to the feature's core. Most of diffuse bands are quite asymmetric
        making the procedure uncertain. Here we estimated the diffuse bands location (its effective center) using the center of gravity;
        This approach is universal and provides simple estimate of a profile asymmetry and allows to detect variability from object to object in rather objective way.
  \item The width of diffuse bands varies form object to object but it can not be correctly measured using FWHM owing to the asymmetric shapes of DIB's profiles.
        The effective width which is a relation of the equivalent width (in m\AA) to its depth (measured in normalized spectra with continuum level equal to one) is free of flaws introduced by FWHM;
\end{itemize}

\section{Results}

As demonstrated by \citep{2015PASP..127..126G}, the galactic orbits of clouds are circular, i.e. thermalized, very
likely due to the collisions of these, geometrically very large objects (see \citet{gm2017}). Such
collisions lead to the mixing of matter, and thus global chemical compositions of individual clouds should not be much different one
from others, at least at similar distances from the galactic center. Typically, observable clouds are revealed by the same set
of atomic spectral features. Mutual intensity ratios of spectral lines/bands are however not identical in all clouds, even of the
same type, e.g. translucent ones, proving the differences in physical parameters, from one object to another. Such differences
of physical parameters of the interstellar translucent clouds, constrain thus possible models of such clouds, reflecting the
variety of physical conditions even in clouds of very similar mass density. Let's emphasize that physical conditions inside
interstellar clouds depend not only on their densities but also on the irradiation by nearby, hot stars.

In our targets  the interstellar atomic/molecular lines do not show evident Doppler splitting of components with comparable intensity, thus DIB profiles are very likely
determined by physical parameters of the intervening clouds only. This is of basic importance as the strong DIBs are apparently
carried by another molecule each (their strength ratios are variable). Having the spectra, shifted to the rest wavelength
velocity frame, we can measure the laboratory wavelengths of the unidentified DIBs. The latter contain, in many cases, some substructures,
following the physical parameters of the intervening clouds.

The measurement of the rest wavelength of diffuse bands requires the shift of the wavelength scale to the "interstellar"\, frame normally made with the aid of
known interstellar atomic or molecular lines. Relatively strong interstellar K{\sc i} at 7699~\AA\ and CH at 4300~\AA\ lines exhibit good or moderate
correlation of the equivalent widths with those of DIBs.
Close up analysis of two "major"\, diffuse bands reveals that EW of so called "narrow"\, diffuse band 5797~\AA\ exhibits good correlation with both K{\sc i} and CH, while
a broader band (5780~\AA) demonstrates poor correlation with CH and better although moderate correlation with potassium.
However, both bands exhibits very high correlation with narrowest known DIB 6196~\AA\ (Fig. 4).
Other studied DIBs -- very broad DIB 6284 and narrow, nearly symmetric DIB 7224 exhibit absolutely similar behavior (Fig. 5).

Hence we conclude that DIB 6196 is a preferable reference point for "interstellar"\, wavelength scale. The carrier of DIB 6196 is not known, i.e. its assigned wavelength
is arbitrary and, in general, may be variable as well. In \cite{2021MNRAS.508.4241K} we demonstrated that variability of DIB 6196 profile occurs mainly in its
upper red part while the profile core remains almost intact, i.e. the almost perfectly symmetrical core of the band profile
makes it possible to unambiguously determine the zero-point of the interstellar (DIB-based) wavelength scale.
Indeed, in Fig. 6 we show DIB 6196's profiles for several objects exhibiting very different effective width  of diffuse bands (Tab. 1).

Indeed, the central wavelength of the full width at half maximum of DIB 6196 in spectra of stars, studied in \cite{2021MNRAS.508.4241K} (except blue-shifted
Sco OB1 objects) lies in the narrow range 6195.9 -- 6196.04 with the average value 6195.96$\pm$ .01 which is almost equal to the assigned wavelength 6195.97 we adopted from
\cite{2000PASP..112..648}. Finally, the advantage of DIB 6196 as a reference for "interstellar"\ wavelength scale is evident in the plots comparing the effective
wavelength of diffuse band with the effective width W$_{eff}$: the outliers belonging to so-called blue-shifted objects (Sco OB1 stars and run-away star HD34078 -- seen in Fig.2 of
\cite{2021MNRAS.508.4241K}) are not outliers anymore. Indeed, the origin of existence of the outliers (Sco OB1 objects)
in Fig.2 of \cite{2021MNRAS.508.4241K}) is the $\sim$7 \kms difference between position of K{\ i} 7699 \AA\ (used there as a reference for conversion of the wavelength scale to the interstellar frame)
and that of diffuse bands (Fig. 7). Obviously, carriers of different diffuse bands share the cloud's volumes between themselves to the greater extent than they do with other interstellar species:
 interstellar potassium, CH, etc.

With increasing the spectral resolution some additional Doppler components in atomic and/or molecular
lines can be detected; in diffuse bands it becomes very difficult and in case of extinction or polarization (caused by dust grains) \textemdash
impossible. Unfortunately, the quantity of high resolution spectra (R$\geq$100,000) is limited.
On the other hand, while even R $\approx$ 90,000 is insufficient for resolving the faint profiles of atomic/molecular interstellar lines,
the resolving power R $\approx$ 48,000 allows us resolve the genuine width of the narrowest known diffuse band at 6196~\AA\ though the faint details of profiles cannot be resolved.
The observed symmetry of DIB 6196 profiles in the core  is very important while we want to use the band as a reference for conversion of the wavelength scale to the interstellar frame.

Close up analysis of Fig. 8 allow us to draw some preliminary conclusions: (i) there is a good correlation between the center of gravity and the broadness (W$_{eff}$) of diffuse bands
7224, 5780 and 5797 (panels {\bf (a, c, d)}). The same correlation for DIB 6284 formally is poor, but this might be an influence of several lines of sight with uncertain measurements (panel {\bf (b)});
(ii) with increasing of the broadness W$_{eff}$, the center of gravity moves to the red, i.e. the feature gets broader to the red-side. Assuming the molecular origin of diffuse bands it can be argued that broader
profiles exhibit more P-branch transitions; (iii) There is a moderate correlation of the center of gravity of DIB 5797 with reddening (r = 0.51, panel {\bf (e)}) and, the correlation
is poor with the equivalent width (r = 0.29, panel {\bf (e)}). It is worth to mention the poor correlation between the width W$_{eff}$ of DIB 5797 with both reddening E$_{B-V}$ and equivalent width EW -
the correlation coefficients are 0.29 and 0.11 respectively (not shown). Similar situation is for other DIBs (not shown); (iv) DIB 5780 is broad in $\sigma$-clouds and this is not due to
the strength of the feature: indeed, EWs of DIB 5780 in the Trapezium objects are quite low while the widths W$_{eff}$ are the highest in the sample.

Once again, DIB 6284 \AA\ is one of the broadest known diffuse bands. It is severely polluted by telluric lines causing the uncertainty of the measurements for this feature and probably causing the lack of correlation
between the center of gravity and the W$_{eff}$ of the band (Fig. 8 {\bf (b)}).
On the other hand, it appears that equivalent width of DIB 6824 is not very sensitive to the measurement imperfection. The band correlates very well with DIBs 5780 and 6196, but not with 5797 (Tab. 2).
This is surprising since both 6196 and 5797 are so-called "narrow" DIBs with very good mutual correlation.

DIB 7224 \AA\ is a narrow and very symmetric feature having almost perfect gaussian-like profile. Telluric contamination introduces additional uncertainty into the measurement results,
but in general there are no serious difficulties in measurements of this band. The DIB exhibits moderate correlation (r = 0.72) between its center of gravity and the width.
It is also seen that $\zeta$ objects exhibit more narrow DIB 7224 profiles while $\sigma$ objects show systematically broader profile (Fig. 8).

In Fig. 8 one can see that DIB 5797 and 7224 profiles are broader but weaker in $\sigma$ objects: in general, red circles are concentrated above the "blue"\, group.
Perhaps, the narrow DIBs 5797 and 7224 get weaker but broader there due to strong UV-flux existing in $\sigma$ clouds which excite the higher transitions of unknown molecular carriers.
This is a bit surprising taking into account commonly accepted association of $\sigma$ clouds with broad diffuse bands.

\begin{figure} 
\centering
\includegraphics[width=8cm]{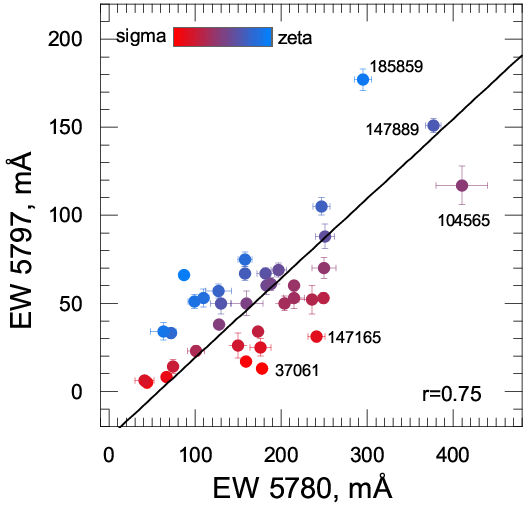}
\caption{A comparison of equivalent widths of "major"\, DIBs. Note the presence of good correlation between these two bands having obviously different origin.
}
\label{d5780d5797}
\end{figure}

\begin{figure} 
\centering
\includegraphics[width=8cm]{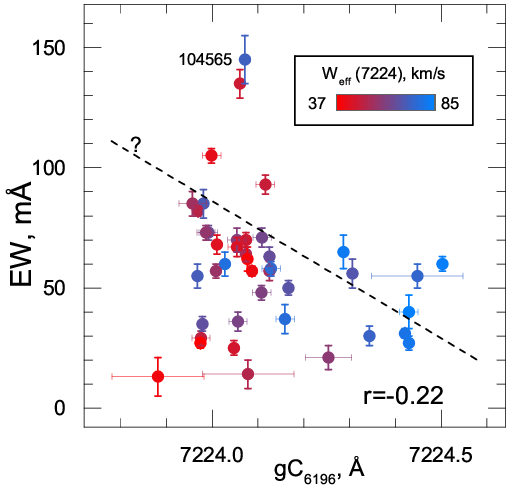}
\caption{Relations between equivalent width of the DIB 7224 profile and its center of gravity.
Note the variability of the DIB7224's width marked with red-blue colors.
}
\label{GC7224}
\end{figure}

\cite{2019MNRAS.486.3537K} demonstrate that in fact there are many intermediate types of clouds and that their spectra are characterized both by different strength ratios of DIBs and of the features carried by interstellar identified molecules and dust grains. We used a relation of the equivalent widths of DIBs 5797 and 5780 as an additional parameter reflecting the physical conditions in the clouds.
Three dimensional relations we offered here (two dimensional scatter plots with color intensity scale for the 3rd parameter - relation of EWs of DIB's 5797 and 5780)
exhibits interesting results: (i) DIB 5780 is weaker in $\zeta$ clouds (Fig. 4, the plot EWs 6196 vs 5780) while DIB 5797 is stronger there (Fig. 4, the plot EWs 6196 vs 5797); (ii) the center of gravity of DIB 5797
is rather red-shifted in $\zeta$ object (Fig. 8), i.e. profile of this DIB is more asymmetric in $\zeta$-clouds. Perhaps it happens owing to the excitation of higher transitions of
the P-branch. Indeed, laboratory spectra show the broadening and red-shift of molecular profiles with increasing temperature.

Finally, let us emphasize that the correlations between interstellar species must be taken into account with precaution because the correlation magnitude
depends on the sample of studied objects. Big samples generally show at least some moderate correlation. A moderate sample of carefully selected single-lined objects are preferable.
Relatively good correlation does not necessarily mean the presence of physical relation between species.
For example, the nominal correlation between major diffuse bands 5780 and 5797 \AA\ measured in our targets (see Fig. 9 and Tab. 2)
is relatively high (the Pearson coefficient r = 0.75) despite of well known different origin of these features. Another interesting example is a peculiar object HD210121 with very strong molecular and atomic lines (including
both K{\sc i} 7699 \AA\ and CH 4300). Despite of relatively good correlation of diffuse bands with these lines (in particular in $\zeta$-clouds) all DIBs in spectra of HD210121 are very weak.

\begin{table}
\centering
\caption{Correlation coefficients between reddening E$_{B-V}$, equivalent width of marked diffuse bands, interstellar lines CH 4300 and K{\sc i} 7769 \AA\, line.
}
\begin{tabular}{rcccccccc}
\hline
        &  CH   & 5780&  5797&  6196& 6284& 7224&K{\sc i}
                                                      &  E$_{B-V}$  \\
\hline
CH      &  1    &0.36 & 0.58 & 0.46 &0.03 &0.07 &0.68 & 0.69 \\
5780    & 0.36  &  1  & 0.75 & 0.92 &0.80 &0.80 &0.52 & 0.69  \\
5797    & 0.58  &0.75 &  1   & 0.90 &0.48 &0.56 &0.62 & 0.74  \\
6196    & 0.46  &0.92 & 0.90 &  1   &0.70 &0.81 &0.67 & 0.65  \\
6284    & 0.03  &0.80 & 0.48 & 0.70 &  1  &0.79 &0.35 & 0.52  \\
7224    & 0.07  &0.80 & 0.56 & 0.81 &0.79 &  1  &0.47 & 0.42  \\
K{\sc i}& 0.68  &0.52 & 0.62 & 0.67 &0.35 &0.47 &  1  & 0.47 \\
\hline
\end{tabular}
\label{pearson}
\end{table}

\begin{table}
\centering
\caption{There are given: the variation range (minimum and maximum) of the center of gravity of diffuse bands and the mean value;
the variation range of W$_{eff}$ of studied diffuse bands and the mean values. All measurements are made for objects with the reddening
E$_{B-V}$ in the range 0.13 -- 1.06 magnitudes.
}
\begin{tabular}{cccccc}
\hline
  \multicolumn{2}{c}{$\lambda_{gc}$(\AA)}
                       &
                                         &  \multicolumn{2}{c}{W$_{eff}$(\kms)} & \\
  min     & max        &  Mean           &     min    &  max       &  Mean          \\
  \hline
 5797.06  & 5797.33    & 5797.17$\pm$0.01&     36     &    55      &  43$\pm$1 \\
 5780.41  & 5781.24    & 5780.68$\pm$0.02&     101    &    140     & 114$\pm$1 \\
 6283.35  & 6284.60    & 6283.94$\pm$0.05&    160     &    264     & 218$\pm$3 \\
 7223.88  & 7224.50    & 7224.13$\pm$0.02&    37      &    85      &  45$\pm$1 \\
\hline
\end{tabular}
\label{amplitude}
\end{table}

\section{Conclusions}

Tables 1, 3 and figures exhibit strong variability of both width and center of gravity of diffuse bands.
Let us emphasize once again that variability of diffuse bands' widths, reported in \cite{2021MNRAS.508.4241K} and in this study, is not caused by the presence of several
clouds with different radial velocities. The same DIB may be narrower or broader depending on the line of sight;
since the depicted targets do not show strong Doppler split in interstellar K{\sc i} line, this must be caused by physical parameters of the intervening objects.
In the range of interstellar reddening 0.13 - 1.06$^{mag}$ we found the following:
\begin{itemize}
\item
   The center of gravity of diffuse band 5797 varies in relatively narrow range 5797.06 - 5797.33 \AA, while its W$_{eff}$ increases in more than one and a half times. This means the lower influence of the
   red wing developed with changing of physical conditions on the entire intensity of the band;
\item
   In contrast, the center of gravity of diffuse band 5780 varies in broad range 5780.41 - 5781.24 \AA, while its W$_{eff}$ increases in a little bit lower extent, i.e. the developed red wings are deep enough
   to move the center of gravity towards the red;
\item
   Broad DIB 6284 exhibits strong variability of both parameters. Once again let's consider these measurement with precaution due to uncertain measurements of this feature;
\item
   DIB 7224 is a rare case of almost gaussian shaped diffuse band. The band demonstrates astonishing variability of the width varying more than 2 times while the center of gravity moves to
   the red-side for $\sim$0.6 \AA\ only. Perhaps, the carriers of this band is very sensitive to variations of physical conditions in the cloud:  both the blue- and red- wings of the profile are extending while
   the profile gets more shallow.
\item
   DIB 5780 is weaker in $\zeta$ clouds (Fig. 4, the plot EWs 6196 vs 5780) while DIB 5797 is stronger there (Fig. 4, the plot EWs 6196 vs 5797)
\item
   The center of gravity of DIB 5797 is rather red-shifted in $\zeta$ object (Fig. 8), i.e. profile of this DIB is more asymmetric in $\zeta$-clouds.
\item
   Trapezium stars ($\sigma$ objects) exhibit broadest diffuse bands in the studied sample.
\end{itemize}

Combining two new approaches (DIB 6196 as a reference and the center of gravity as a position mark) we found that blue- and red- shifted objects (Sco OB1 objects, HD 34078, Trapezium objects)
see  e.g. \cite{2021MNRAS.508.4241K}) are not outliers in the relation "DIB position - its width" but exhibit a gradual relation. However, in some cases the red-shift caused by the broadening of the profile
to its red-side is accompanied by complete displacement of the profile - see DIB 5780 profiles in Fig. 8). The remaining question is whether this displacement is caused by (i) the Doppler effect taking place if
the carriers of studied DIBs and those of DIB 6196 (serving as a reference for the interstellar wavelength scale) are occupying different cloud volumes or we are observing profile variations caused by physical effects, e.g. higher/lower temperature, variations of the isotopes abundances  (an example is HD\,37023's profile of DIB 5780).

Diffuse band 6284 is a feature difficult for analysis due to its broadness and severe pollution by telluric lines (the O$_2$ $\alpha$ band) making its profile restoration too uncertain for solid conclusions.

Diffuse band 7224 shows rather poor but probably existing anti-correlation between the center of gravity and the equivalent width (Fig. 10). Figure 10 demonstrates that
stronger features are more narrow but more deep. In combination with relation between the center of gravity and the width shown in Fig. 8 we conclude that DIB 7224 of
 $\zeta$-objects concentrated in the lower part of the plot  are more narrow and weak in $\zeta$-clouds and more broad and strong than in $\sigma$-clouds.
As we stated in \citet{2021MNRAS.508.4241K}, the observed broadening of diffuse bands can be used as an additional criterion for grouping these features into families, presumably formed by the same molecule.
Would be interesting to find other "narrow"\, diffuse bands belonging to the $\sigma$ family.

The next step of the great importance is to find possible connection between the magnitudes of the broadening and red-shift of diffuse band profiles and
variation of physical parameters in the cloud's volumes where the DIBs carriers are located. Precisely estimated rotational temperature of e.g. interstellar C$_2$ molecule (well correlating with many diffuse bands)
in the single-cloud lines of sight is more than welcome. Very limited number of ultra-high signal-to-noise ratio and high spectral resolution spectra remains a problem.

\begin{acknowledgments}

This research has made use of the services of the ESO Science Archive Facility and the SIMBAD database, operated at CDS, Strasbourg, France (Wenger et al. 2000).
\end{acknowledgments}

\section*{FUNDING}
GAG made data processing and analysis with the aid of instruments developed at Special Astrophysical Observatory with the financial support of
the Ministry of Science and Higher Education of the Russian Federation under the grant 075-15-2020-780 (N13.1902.21.0039)

\section*{CONFLICT OF INTEREST}
The authors declare no conflicts of interest.

\bibliographystyle{aspb1}
\bibliography{DIBwave}

\begin{thebibliography}{25}
\providecommand{\natexlab}[1]{#1}

\bibitem[{Bernstein} et al.(2003)]{Ber03} R. {Bernstein}, S.~A. {Shectman}, S.~M. {Gunnels} et al., SPIE, \textbf{4841}, 1694 (2003).
\bibitem[{Dekker} et al.(2000)]{Dekal00} H. {Dekker}, S. {D'Odorico}, A. {Kaufer} et al., SPIE, \textbf{4008}, 534 (2000).
\bibitem[{Fan} et al.(2019)]{Fan 2019} H. {Fan}, L.~M. {Hobbs}, J.~A. {Dahlstrom} et al., \apj, \textbf{878}, 151 (2019).
\bibitem[{Galazutdinov} et al.(2000)]{2000PASP..112..648} G.~A. {Galazutdinov}, F.~A. {Musaev}, J. {Kre{\l}owski}  et al., \pasp, \textbf{112}, 648 (2000).
\bibitem[{Galazutdinov}  et~al. (2006)]{G06} G.~A. {Galazutdinov}, G. {Manic{\`o}}, J. {Kre{\l}owski}, \mnras, \textbf{366}, 1075 (2006).
\bibitem[{Galazutdinov} et al.(2008)]{2008PASP..120..178G} G.~A. {Galazutdinov}, G. {LoCurto}, I. {Han}, J. {Kre{\l}owski}, \pasp, \textbf{120}, 178 (2008).
\bibitem[{Galazutdinov}, {LoCurto} \& {Kre{\l}owski}(2008)]{2008ApJ...682.1076G} G.~A. {Galazutdinov}, G. {LoCurto}, J. {Kre{\l}owski}, \apj, \textbf{682}, 1076 (2008).
\bibitem[{Galazutdinov} et al.(2015)]{2015PASP..127..126G} G.~A. {Galazutdinov}, A. {Strobel}, F.~A. {Musaev}  et al., \pasp, \textbf{127}, 126 (2015).
\bibitem[{Galazutdinov} et al.(2020)]{2020AJ....159..113G} G. {Galazutdinov}, A. {Bondar}, B.-C. {Lee} et al., \aj, 159, \textbf{113} (2020).
\bibitem[{Galazutdinov} (2022)]{dech} G.~A. {Galazutdinov},  Astrophysical Bulletin, \textbf{77}, 519 (2022).
\bibitem[{Gnaci{\'n}ski} \& {M{\l}ynik} (2017)]{gm2017} P. {Gnaci{\'n}ski} \& T. {M{\l}ynik}, \pasp, \textbf{129}, 044101 (2017).
\bibitem[{Hartmann} (1904)]{Hartmann1904} J. {Hartmann}, \apj, \textbf{19}, 268 (1904).
\bibitem[{Heger} (1922)]{Heger1922} M.~L. {Heger}, Lick Observatory Bulletin, \textbf{10}, 146 (1922).
\bibitem[{Kaufer} et al.(1999)]{Ketal99} A. {Kaufer}, O. {Stahl}, S. {Tubbesing} et al., The Messenger, \textbf{95}, 8 (1999).
\bibitem[{Ka{\'z}mierczak} et al.(2009)]{K2009} M. {Ka{\'z}mierczak}, P. {Gnaci{\'n}ski}, M.~R. {Schmidt}, G. {Galazutdinov}, A. {Bondar}, J. {Kre{\l}owski}, \aap, \textbf{498}, 785 (2009).
\bibitem[{Kerr} et al.(1998)]{Kerr1998} T.~H. {Kerr}, R.~E. {Hibbins}, S.~J. {Fossey}  et al., \apj, \textbf{495}, 941 (1998).
\bibitem[{Kim} et al.(2007)]{kimetal2007} K.-M. {Kim}, I. {Han}, G.~G. {Valyavin} et al., \pasp, \textbf{119}, 1052 (2007).
\bibitem[{Kre{\l}owski} \& {Walker} (1987)]{KW87} J. {Kre{\l}owski} \&  G.~A.~H. {Walker}, \apj, \textbf{312}, 860 (1987).
\bibitem[{Kre{\l}owski} \& {Westerlund} (1988)]{KW88} J. {Kre{\l}owski} \& B.~E.{Westerlund},  \aap, \textbf{190}, 339 (1988).
\bibitem[{Kre{\l}owski} \& {Greenberg} (1999)]{KG99} J. {Kre{\l}owski} \&  J.M. {Greenberg}, \aap,  \textbf{346}, 199 (1999).
\bibitem[{Kre{\l}owski}  et~al. (2015)]{2015MNRAS.451.3210K} J. {Kre{\l}owski}, G. A. {Galazutdinov}, G. {Mulas}, M. {Maszewska} \& C. {Cecchi-Pestellini}, \mnras,  \textbf{451}, 3210 (2015).
\bibitem[{Kre{\l}owski} et al.(2019)]{2019MNRAS.486.3537K} J. {Kre{\l}owski}, G. A. {Galazutdinov} \& A. {Bondar}, \mnras, \textbf{486}, 3537 (2019).
\bibitem[{Kre{\l}owski} et al.(2021)]{2021MNRAS.508.4241K} J. {Kre{\l}owski}, G. A. {Galazutdinov}, P. {Gnaci{\'n}ski} et al., \mnras, \textbf{508}, 4241 (2021).
\bibitem[{Motylewski} et al.(2000)]{motyl2000} T. {Motylewski}, H. {Linnartz}, O. {Vaizert} et al., \apj, \textbf{531}, 312 (2000).
\bibitem[{Pepe} et al.(2014)]{PMC14} F. {Pepe}, P. {Molaro}, S. {Cristiani} et al., 2014, Astronomische Nachrichten, \textbf{335}, 8 (2014).
\bibitem[{Sarre} et al.(1995)]{Sarre1995} P.~J. {Sarre}, J.~R. {Miles}, T.~H. {Kerr} et al., \mnras, \textbf{277}, L41 (1995).
\bibitem[{Salama} et al.(1999)]{Salama1999} F. {Salama}, G.~A. {Galazutdinov}, J. {Kre{\l}owski} et al., \apj, \textbf{526}, 265 (1999).
\bibitem[{Siebenmorgen} et al.(2020)]{Siebenmorgen2020} R. {Siebenmorgen}, J. {Kre{\l}owski}, J. {Smoker} et al., 2020, \aap, \textbf{641}, A35 (2020).
\bibitem[{Wenger} et al.(2000)]{Wenger} M. {Wenger}, F. {Ochsenbein},  D. {Egret} et al., \aaps, \textbf{143}, 9 (2000).

\end{thebibliography}


\end{document}